\documentclass[runningheads]{llncs}
\usepackage[T1]{fontenc}
\newcommand{\mycomment}[1]{}

\usepackage{amsmath}
\usepackage{amsfonts}
\usepackage{algorithmicx}
\usepackage{balance}
\usepackage{float}
\usepackage{tcolorbox}
\usepackage{hyperref}
\usepackage{bbding}
\usepackage{nopageno}
\usepackage{algorithm}
\usepackage[noend]{algpseudocode}
\usepackage{graphicx}
\usepackage{textcomp}
\usepackage{multirow}
\usepackage{enumitem}
\usepackage{pifont}
\usepackage{xcolor}
\usepackage{bm}
\usepackage{tikz}
\usetikzlibrary{shapes.geometric, arrows}
\usepackage[T1]{fontenc}
\usepackage{array}
\usepackage{ragged2e}
\usepackage{threeparttable}
\newcolumntype{M}[1]{>{\centering\arraybackslash}m{#1}}

\newcommand\aamtl{\textsf{DA-MTL}}
\newcommand\multitude{MULTITuDE}
\newcommand\red{\textcolor{red}}
\newcommand\blue{\textcolor{blue}}
\newcommand\myplus{\begin{tiny}\raisebox{+0.65ex}{+}\end{tiny}}
\newcommand\myminus{\begin{tiny}\raisebox{+0.65ex}{-}\end{tiny}}

\newcolumntype{P}[1]{>{\centering\arraybackslash}p{#1}}

\begin{document}

\title{Two Birds with One Stone: Multi-Task Detection and Attribution of LLM-Generated Text}
\titlerunning{Multi-Task Detection and Attribution of LLM-Generated Text}

\author{
  Zixin Rao\inst{1}\textsuperscript{*}   \and   
  Youssef Mohamed\inst{2}\textsuperscript{*}  \and
  Shang Liu\inst{3} \and
  Zeyan Liu\inst{3}\textsuperscript{**}
}

\institute{
  University of Georgia \\
  \email{zr04546@uga.edu}
  \and
  Egypt-Japan University of Science and Technology \\
  \email{youssef.khalil@ejust.edu.eg}
  \and
  University of Louisville \\
  \email{\{shang.liu, zeyan.liu\}@louisville.edu}
}

\maketitle

\renewcommand{\thefootnote}{\fnsymbol{footnote}}
\footnotetext[1]{These authors contributed equally to this work, which was conducted during their internship at the University of Louisville.}
\footnotetext[2]{Corresponding author.}

\begin{abstract}
Large Language Models (LLMs), such as GPT-4 and Llama, have demonstrated remarkable abilities in generating natural language. However, they also pose security and integrity challenges. Existing countermeasures primarily focus on distinguishing AI-generated content from human-written text, with most solutions tailored for English. Meanwhile, authorship attribution--determining which specific LLM produced a given text--has received comparatively little attention despite its importance in forensic analysis. In this paper, we present \aamtl, a multi-task learning framework that simultaneously addresses both text detection and authorship attribution. We evaluate \aamtl~on nine datasets and four backbone models, demonstrating its strong performance across multiple languages and LLM sources. Our framework captures each task’s unique characteristics and shares insights between them, which boosts performance in both tasks. Additionally, we conduct a thorough analysis of cross-modal and cross-lingual patterns and assess the framework’s robustness against adversarial obfuscation techniques. Our findings offer valuable insights into LLM behavior and the generalization of both detection and authorship attribution. Our source code is available at \url{https://github.com/youssefkhalil320/MTL_training_two_birds_with_one_stone}.

\end{abstract}

%
%
%
\raggedbottom

\section{Introduction} \label{sec:intro}

Neural language generation has revolutionized natural language processing, enabling the creation of highly coherent, contextually relevant, and human-like text \cite{wei2022emergent}. Powered by advances in Large Language Models (LLMs), such as GPT-4 \cite{achiam2023gpt} and Llama \cite{touvron2023llama1,touvron2023llama2}, these models have achieved extraordinary performance across diverse applications, including education \cite{yan2024practical}, medicine\cite{thirunavukarasu2023large}, software engineering\cite{hou2023large}, creative industries \cite{gruda2024three}, and so on. However, the proliferation of LLM-generated content has also raised widespread concerns related to integrity, security, creativity, and ethics \cite{yao2024survey,anders2023using,derner2023beyond,lu2022effects}. Meanwhile, the accessibility and ease of use of commercial tools such as ChatGPT \cite{chatgpt} and Google Gemini \cite{gemini} have further magnified these risks. For example, the proportion of articles generated by LLMs on misinformation websites has grown rapidly \cite{hanley2024machine}, and as of December 2023, 26.1\% of arXiv submissions included ChatGPT-generated content \cite{liu2023detectability}. In addition, threat actors have exploited LLMs for malicious purposes such as phishing and website hacking \cite{roy2023generating,grbic2023social,fang2024llm}. These trends highlight the urgent need for effective mechanisms to mitigate the social impacts of LLM-generated content.

Detecting LLM-generated text has emerged as a critical and commonly used solution to address these challenges, both in research and in practice \cite{gptzero,tang2024science}. Existing studies primarily focus on identifying whether a given text is human-written or AI-generated, with most studies centered on English or closely related languages and GPT models \cite{zellers2019defending,kushnareva2021artificial,guo2023close,liu2023argugpt,tu2023chatlog,liu2023detectability}. However, these approaches leave gaps in understanding how detection generalizes across diverse languages and LLM architectures. Recent works, such as \multitude \cite{macko2023multitude} and M4 \cite{wang2024m4,wang2024m4gt}, have started to explore multi-lingual and multi-modal detection, but questions regarding cross-linguistic and cross-modal generalization remain unanswered.

Authorship attribution, which identifies the specific AI model responsible for generating a piece of text, is an equally important but underexplored challenge. With the rapid proliferation of LLMs-hundreds being developed each week-and the growing sophistication of generated text \cite{gao2023origin}, accurately attributing text to its source model is becoming increasingly difficult. At the same time, this task is critical for applications such as intellectual property protection \cite{huang2024authorship}. Unlike detection, attribution requires a more nuanced analysis of characteristics. This makes it a technically complex but essential problem to regulate and understand AI-generated content. However, most existing studies focus only on binary classification between human-written and machine-generated text. In addition, many recent studies on authorship attribution overlook human-written text or simply treat it on par with individual LLMs \cite{uchendu2020authorship,munir2021through,chen2023token,sotofew}.

In this paper, we introduce a new LLM forensics framework called \aamtl, which simultaneously tackles text detection and authorship attribution using multi-task learning. Our approach uses shared representations to solve two tasks: (1) distinguishing between human-written and machine-generated text, and (2) identifying the specific source LLM of machine-generated text. We evaluate the framework on multiple backbone models, datasets, and languages, demonstrating its superior performance. Notably, our framework improves both tasks. Based on our accurate models, we explore cross-modal similarities, investigating how shared patterns among LLMs influence their proximity in feature space and their attribution. We also examine how LLMs differ using stylometric analysis and how our forensics models trained in one language family generalize to others. Finally, we test the robustness of the proposed framework against obfuscation techniques, showing its ability to handle content post-processed by adversaries.

Specifically, we aim to answer the following research questions:
\begin{itemize}
\item \textbf{RQ1}: Can multi-task learning improve text detection and attribution performance compared to single-task baselines?
\item \textbf{RQ2}: Which LLMs are similar, and how do these similarities impact attribution performance?
\item \textbf{RQ3}: How does \aamtl~generalize across languages?
\end{itemize}

Our main contributions include:
\begin{enumerate}
\item We present a multi-task learning framework, \aamtl, to jointly solve text detection and attribution, which shows outstanding performance.
\item We comprehensively analyze cross-modal and cross-lingual generalization, which provides deep insights into LLM proximity and behaviors.
\item {We demonstrate the robustness of \aamtl~against adversarial obfuscation techniques, showcasing its effectiveness in complex real-world scenarios.}
\end{enumerate}

The rest of the paper is organized as follows: Section~\ref{sec:related} reviews related work. Section~\ref{sec:method} formalizes the problem and presents our \aamtl~framework. Section~\ref{sec:exp} shows experimental results and analysis, followed by our conclusions in Section~\ref{sec:conclusion}.
\section{Related Work}  \label{sec:related}
\noindent\textbf{Machine-Generated Text Detection:}
The growing capabilities of language models to generate realistic text have led to a significant increase in research efforts for machine-generated text (MGT) detection. Typically framed as binary classification, MGT detection distinguishes between machine-generated and human-written text. Early methods relied on linguistic and statistical discrepancies, employing classical feature-based classifiers \cite{gehrmann2019gltr,mitchell2023detectgpt,su2023detectllm}. Recent advances have embraced fine-tuned transformer models such as RoBERTa, achieving performance improvements in identifying subtle stylistic and structural differences \cite{solaiman2019release,guo2023close,liu2023detectability,liu2023argugpt}. Furthermore, the introduction of comprehensive datasets and corpora, such as those featuring AI-generated responses to consistent instructions, has significantly advanced this field by offering benchmarks \cite{liu2023argugpt,liu2023detectability,macko2023multitude,wang2024m4,guo2023close,he}.

\vspace{1mm}\noindent\textbf{Authorship Attribution:}
Authorship attribution is a traditional problem in NLP, which identifies the author of a text based on linguistic and stylistic cues \cite{solorio2011modality}. In the context of LLMs, attributing machine-generated texts is particularly important to ensure proprietary and ethical responsibility when using AI-generated content \cite{uchendu2020authorship}. LLMs often embed unique stylistic signatures, which have been utilized by many methods \cite{uchendu2020authorship,venkatraman-etal-2024-gpt,hu2020deepstyle}. Furthermore, advanced models such as XLM-R and RoBERTa also demonstrated effectiveness \cite{wang2018glue}. However, there are still challenges like cross-domain attribution and complex datasets, emphasizing the need for continued advances in this area \cite{stamatatos2009survey,kestemont2018overview}. In adversarial scenarios, it is also crucial that attribution techniques be robust against intentional obfuscation techniques employed to conceal or mask authorship \cite{keswani2016author,almishari2014fighting,macko2024authorship,mahmood2019mutantx}.

\vspace{1mm}\noindent\textbf{Multi-task Learning for NLP:}
Multi-task learning (MTL) has emerged as a foundational approach to improve performance across related tasks. Initially introduced by Caruana \cite{caruana1997multitask}, MTL has been shown to improve generalization and reduce overfitting by regularizing models through shared knowledge \cite{ruder2019latent,sener2018multi}. In NLP, early applications demonstrated its utility for tasks such as part-of-speech tagging and sentiment analysis \cite{ando2005framework,liu2019multi}, while language models like T5 \cite{raffel2020exploring} successfully framed multiple NLP tasks as text generation problems. Cross-lingual tasks have also benefited from MTL \cite{taslimipoor2019cross}, while optimization techniques have been developed to address cross-task gradient conflicts \cite{yu2020gradient}. In this paper, we extend MTL by applying it to MGT detection and attribution.

\section{The \aamtl~Framework}  \label{sec:method}
\subsection{Problem Definition and Assumptions} \label{sec:definition}
We specify two tasks in our paper: \textbf{LLM Detection (Task 1 or $\mathcal{T}_{D}$)} and \textbf{LLM Attribution (Task 2 or $\mathcal{T}_{A}$)}. Specifically, Task 1 determines whether a given text is human-written or generated by an LLM. Task 2 identifies the specific LLM responsible for generating the text, which involves multiple classes.


Formally, given a dataset  $\mathcal{D} = \{(x_i, y_{\mathcal{T}_{D},i}, y_{\mathcal{T}_{A},i})\}^N_{i=1}$, where $x_i$ is a piece of text, $y_{\mathcal{T}_{D},i} \in \{0, 1\}$ is the binary label for detection, and $y_{\mathcal{T}_{A},i} \in \{1, 2, \ldots, C\}$ is the multi-class label for attribution, and $N$ is the total number of samples. We assume that the data are sampled i.i.d. from a joint distribution $\mathcal{P}(x, y_{\mathcal{T}_{D}}, y_{\mathcal{T}_{A}})$, or $\mathcal{P}$ in short. Our goal is to optimize a multi-task classifier with parameters $\theta$ by maximizing the likelihood for both $\mathcal{T}_{A}$ and $\mathcal{T}_{D}$, which can be approximated by joint optimization as:

\begin{equation}
\min_{\theta} \mathcal{L} = \mathbb{E}_{x \sim \mathcal{P}(x)} \left[ \sum_{\mathcal{T} \in \{\mathcal{T}_{D}, \mathcal{T}_{A}\}} \mathbb{E}_{y_\mathcal{T} \sim \mathcal{P}(y_\mathcal{T} \mid x)} \big[ \mathcal{L}_\mathcal{T}(f_\mathcal{T}(x; \theta), y_\mathcal{T}) \big] \right]
\end{equation}

\noindent where $\mathcal{L}_{D}$ is the binary cross-entropy loss for detection and $\mathcal{L}_{A}$ is the categorical cross-entropy loss for attribution.

Our primary objective is to determine whether a given text was written by human beings or by one of several possible LLM candidates. We assume a \textbf{black-box} defender, which is widely adopted for AI-generated content detection \cite{gptzero,copyleaks,hive}. Specifically, we have no access to the LLM's internal weights, architecture, or gradients. Instead, our analysis relies solely on the model's output. This assumption is necessary and practical for two key reasons: (1) Many leading commercial LLMs, including OpenAI's GPT-4 \cite{achiam2023gpt,chatgpt}, Google's Gemini \cite{gemini}, and Anthropic's Claude \cite{claude}, are proprietary and closed-source. Meanwhile, these models also have the greatest real-world impact with the largest user base. (2) Security-critical AI-generated content may appear in public domains without clear authorship, such as social media posts or research articles. In such cases, a black-box approach is the most effective and feasible way to analyze and verify content in real-world forensic scenarios.

\subsection{Key Observations and Motivation} \label{sec:motivation}
\begin{figure}[tbhp] 
    \centering
    \includegraphics[clip, trim=60 15 15 15, width=0.98\textwidth]{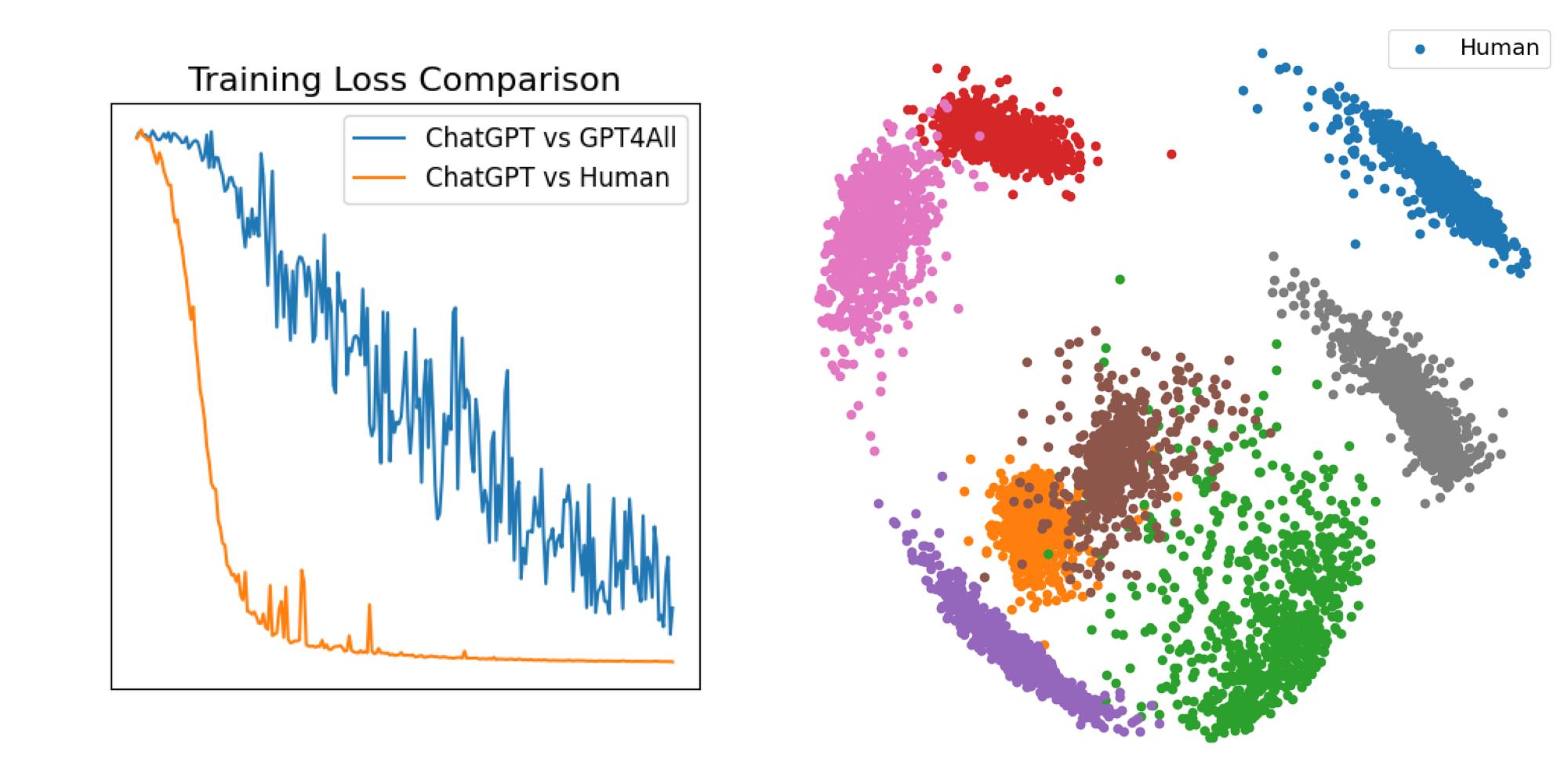}
    \vspace{-2mm}  
    \caption{Left: The training losses for ChatGPT-Human (the orange line) and ChatGPT-GPT4All (the blue line). The training losses decrease much more rapidly when distinguishing human data. Right: PCA visualization of the features, showing a distinct and separate cluster for human data, far from the clusters of LLMs.} 
    \label{fig:motivation}
\end{figure}

LLM detection and attribution inherently involve differences in granularity. Although optimized to mimic human distribution, LLM outputs typically present distinct differences from natural human data \cite{guo2023close}, making Task 1 relatively coarse-grained and straightforward. In contrast, Task 2 could be subtle (e.g., ChatGPT vs. GPT4All), as different LLMs sometimes share architectural similarities and stylistic traits \cite{naveed2023comprehensive}. Toward a better understanding, we showcased two key observations on the Essay dataset in MGTBench \cite{he}:

\begin{itemize}
\item \textbf{Observation I: Task 1 is easier.} We trained two binary classifiers for simplified Task 1 and 2. The training loss curves, shown in Figure~\ref{fig:motivation}, reveal a significant gap in dynamics. The loss for the human versus ChatGPT classifier decreases rapidly and stabilizes early, indicating that this task is much easier to learn. In contrast, the loss for the GPT4All versus ChatGPT classifier decreases more slowly and exhibits greater variability, indicating increased difficulty.

\item \textbf{Observation II: Human data attribution is distinct.} We trained a single multi-class classifier for text generated by human beings as well as different LLMs. We visualized the feature representations using PCA. As shown in Figure~\ref{fig:motivation}, the human data form a distinct cluster, while many LLMs' representations overlap significantly. This suggests that a single classifier might struggle to reconcile both coarse and fine-grained distinctions.
\end{itemize}

One alternative solution is to remove human data from attribution and build a separate, dedicated classifier for detection. However, this approach has two limitations: Firstly, in real-world applications, attributing the authorship of unlabeled text to either a human or one of the LLMs in a unified way is more interesting and practically useful. Secondly, two tasks can hypothetically benefit from each other: (1) understanding subtle stylistic and distributional differences among LLMs enriches the representation, and (2) human-generated content can be a regularizer for attribution, helping with capturing broader linguistic features. We validate these hypotheses in Sec~\ref{sec:crosslingual} by showing how joint training improves both tasks.

These observations motivate us to decompose tasks, where we explicitly retain human-written text in the label space and design a framework to jointly handle detection and attribution. Our goal is to achieve a dual benefit by taking advantage of both human and MGT data to maximize overall performance.

\subsection{Methodology} \label{sec:methodology}
Our framework consists of a shared encoder and two task-specific heads for detection and attribution. Given an input text $x$, the shared encoder produces a $d$-dimensional feature vector $h$.

\begin{equation}
h = \text{SharedEncoder}(x, a)
\end{equation}

Next, each task-specific head applies a linear transformation to $h$, yielding logits for the respective classification tasks. For detection (Task 1), we have a binary classification head, and for attribution (Task 2), we have a $C$-ary classification head:


\begin{equation}
p_{D}(x) = \sigma(W_{D} h + b_{D}), \quad
p_{A,j}(x) = \frac{e^{W_{A,j} h + b_{A,j}}}{\sum_{k=1}^C e^{W_{A,k} h + b_{A,k}}}
\end{equation}

\noindent where $W_{D} \in \mathbb{R}^{1 \times d}$, $W_{A} \in \mathbb{R}^{C \times d}$ are the weights for detection and attribution heads, and $b_D, b_A \in \mathbb{R}$ are the biases. $C$ is the number of LLM candidates. $\sigma(\cdot)$ denotes the sigmoid function. The cross-entropy losses for Task 1 and 2 can be computed as:

\begin{equation}
\begin{aligned}
\mathcal{L}_{D} = - \frac{1}{N} \sum_{i=1}^N \Big[ y_{\mathcal{T}_{D},i} \log(p_{D}(x_i)) + (1 - y_{\mathcal{T}_{D},i}) \log(1 - p_{D}(x_i)) \Big] \\
\mathcal{L}_{A} = - \frac{1}{N} \sum_{i=1}^N \sum_{j=1}^C y_{\mathcal{T}_{A},i,j} \log(p_{A,j}(x_i))
\end{aligned}
\end{equation}

During optimization, gradients for each task are computed independently as $\nabla_{\theta} \mathcal{L}_{D}$ and $\nabla_{\theta} \mathcal{L}_{A}$ and then combined to update the model parameters:

\begin{equation}
\theta \leftarrow \theta - \eta \left( \lambda_{D} \cdot \nabla_{\theta} \mathcal{L}_{D} + \lambda_{A} \cdot \nabla_{\theta} \mathcal{L}_{A} \right)
\end{equation}

\noindent where $\eta$ is the learning rate. In our paper, we set $\lambda_{D}=\lambda_{A}=0.5$.

\vspace{1mm}\noindent\textbf{Design Choice.} Conflicting optimization objectives of different tasks is a common issue in multi-task learning, which can hinder performance. However, we found that adding PCGrad \cite{yu2020gradient}, which projects $\mathcal{T}_{D}$'s gradients onto the normal plane of $\mathcal{T}_{A}$'s gradients, only yields minimal improvements but significantly increases the time and GPU memory costs. Therefore, we choose not to use PCGrad.

\section{Experiments}  \label{sec:exp}

We implemented \aamtl~with PyTorch 2.10.2, Transformers 4.44.2, and Python 3.10.2 on Ubuntu 22.04.3. The experiments are conducted using an Nvidia L4 GPU and an Intel Xeon CPU. Pre-trained model parameters are obtained from Huggingface. For training, we adopt an AdamW optimizer with an initial learning rate of 1e-5, a maximum of 30 epochs, and an early-stopping strategy to stop when the validation F1 score does not increase for six epochs. Detailed experiment settings are listed in Table~\ref{tab:experiment_settings}.
\begin{table}[!htbp]
\centering
\caption{Experiment Settings and Details.} 
\label{tab:experiment_settings}
\setlength{\tabcolsep}{3.5 pt}
\renewcommand{\arraystretch}{1.2}  
\begin{tabular}{p{3cm}|p{7cm}}
\hline
\textbf{Setting} & \textbf{Details} \\
\hline
Environment & Ubuntu 22.04.3, Python 3.10.2, PyTorch v2.4.1+cu121, Transformers=4.44.2 \\ 
\hline
Hardware & NVIDIA L4 GPU \& Intel Xeon CPU @ 2.20GHz\\ 
\hline
Optimizer & AdamW \\ 
\hline
Scheduler & Linear \\ 
\hline
Train/Val Split & 80/20 (if not specified by the dataset) \\ 
\hline
Hyperparameters & $\text{LR}=1\times10^{-5}$, Batch Size=8 to 32, Dropout=0.5, Max Epochs=30, Weight Decay=0.01, Random Seed=42 \\
\hline
Early Stopping & Val F1-score does not increase for $6$ epochs \\
\hline
\end{tabular}
\end{table}

\subsection{Basic Results} \label{sec:baseline}
We first evaluate \aamtl~on mono-lingual tasks. We select \textbf{English}, as it is the most widely studied language in NLP. Specifically, we used nine different datasets from three benchmarks: 
\begin{itemize}
\item \textbf{MGTBench:} A benchmark for detecting AI-generated text across multiple genres. It includes three domains: \textbf{Essay}, \textbf{WritingPrompt (WP)}, and \textbf{Reuters}, which cover diverse writing styles, from formal essays to creative writing and news articles \cite{he}. The dataset features text from six major LLMs: ChatGPT-turbo, ChatGLM, Dolly, GPT4All, StableLM, and Claude, with each domain containing 1,000 samples per generator.
\item \textbf{\multitude:} A large-scale dataset comprising both AI-generated and human-written news articles \cite{macko2023multitude}. It includes 74,081 samples spanning 11 languages, generated by eight multilingual LLMs, namely GPT-3, ChatGPT, GPT-4, Llama-65B, Alpaca-LoRa-30B, Vicuna-13B, OPT-66B, and OPT-IML-Max-1.3B.
\item \textbf{M4:} A benchmark for multi-generator, multi-domain, and multi-source AI-generated text detection. It includes model outputs across various domains such as \textbf{Wikipedia}, \textbf{Reddit}, \textbf{WikiHow}, \textbf{PeerRead}, and \textbf{arXiv} paper abstracts. The dataset features text from six LLMs: GPT-4, ChatGPT, GPT-3.5, Cohere, Dolly-v2, and BLOOMz-176B, which contains 2,344 to 3,000 samples for each generator and domain.
\end{itemize}

The selected datasets comprehensively span a wide range of doamins, including academic writing, news articles, open-ended question-and-answer tasks, user reviews, and instructional materials. They also represent text generated by two categories of LLMs: (1) \textbf{Conversational Agents}: MGTBench and M4 feature responses from application-level LLMs such as ChatGPT, Claude, ChatGLM, and GPT4All, which are designed for real-world deployment via user interface or APIs. These models are more likely to be used in practical settings, raising security concerns; and (2) \textbf{Foundational NLP Models}: MULTITuDE and M4 include text data composed by cutting-edge open-source language models like LLaMA, Vicuna, OPT, and Alpaca. These models are widely employed for research and development, offering a flexible foundation for advanced NLP tasks.

We compare \aamtl~with two categories of baseline models that are commonly adopted in the literature: (1) \textbf{Feature-based ML Classifiers:} These classifiers rely on traditional text embeddings, such as TF-IDF and Bag-of-Words. They have been widely used for various text classification tasks, including MGT detection \cite{ippolito2020automatic,jawahar2020automatic,kushnareva2021artificial,liu2023detectability,kumarage2023neural,zhang2024enhancing,krawczyk2024towards}. We evaluated four machine learning models: Multinomial Naive Bayes (MNB), Linear Regression (LR), Support Vector Machines (SVM), and Random Forests (RF). (2) \textbf{Fine-tuned Language Models} These models are fully fine-tuned on each dataset. We select BERT \cite{kenton2019bert,zellers2019defending,ippolito2020automatic,uchendu2021turingbench,kushnareva2021artificial}, RoBERTa \cite{uchendu2021turingbench,kumarage2023neural,guo2023close,liu2023argugpt,macko2023multitude,wang2024m4}, DistilBERT\cite{sanh2019distilbert,he}, and GPT-2 \cite{zellers2019defending,liu2023detectability}. We also include mBERT \cite{kenton2019bert} and XLM-R \cite{conneau2020unsupervised}, which are designed for multi-lingual problems. In total, we evaluate 10 baseline models across nine datasets, resulting in 90 trained baseline classifiers. For these baseline models, human-written text is treated equally as a separate category alongside LLM-generated content, framing the problem as a single multi-class classification task.

To answer \textbf{RQ1}, we choose four backbone models for comparison with varying parameter sizes, ranging from 67 million to 278 million parameters: DistilBERT, mBERT, XLM-R, and RoBERTa. These models are trained using \aamtl~under the same settings as the baselines. This helps evaluate the general applicability of our \aamtl~framework.

As shown in Table~\ref{table:basic}, we have the following findings: (1) Feature-based machine learning classifiers perform poorly. Especially on the \multitude~dataset, they achieve an average F1 score of just 46.8\%. In contrast, fine-tuned language models perform significantly better, with F1 scores exceeding 86\% on all datasets except \multitude. However, even for \multitude, their performance remains unsatisfactory, with an average F1 score of 71\%. (2) All four \aamtl~models outperform their corresponding baselines trained without \aamtl. On the Essay, WP, and Reuters datasets, \aamtl~yields average F1 score improvements of 2.7\%, 2.1\%, and 1.2\%, respectively. The largest improvement is observed on \multitude, where \aamtl~boosts performance by an average of 4.2\%. It also enhances F1 scores on M4, improving by 3.1\%, 0.5\%, 1.7\%, 1.2\%, and 1.7\% for arXiv, PeerRead, Reddit, Wikihow, and Wikipedia, respectively. (3) We find that \aamtl~is particularly effective on more challenging datasets (such as Essay, \multitude, and arXiv) and with less optimal backbone models (such as XLM-R). The most substantial single improvement occurs with XLM-R on \multitude, where \aamtl~increases the F1 score by 6.5\%. Across all nine datasets, \aamtl~improves F1 scores by an average of 1.9\% for DistilBERT, 2.0\% for mBERT, 2.7\% for XLM-R, and 1.6\% for RoBERTa. This aligns with our expectation that multi-task learning mitigates the dominance of either task, which is more pronounced in overfitted models with larger parameter sizes. (4) Among all classifiers, RoBERTa with \aamtl~achieves the best performance on most of the datasets, except on \multitude, PeerRead and WikiHow, where XLM-R performs best.

\begin{tcolorbox}[title=Takeaway I]
\aamtl~is generally applicable and effective across different LLMs and domains. 
\end{tcolorbox}

\vspace{0mm}\noindent\textbf{Efficiency:} \aamtl~is notably lightweight, adding only a few thousand parameters compared to the hundreds of millions in the encoder. Although training time per epoch is roughly doubled (because the model is invoked twice), \aamtl~achieves good performance in just a few epochs. For instance, DistilBERT reaches its optimal performance in 11 epochs, but with \aamtl, it achieves the same in just 3 epochs. XLM-R with \aamtl~takes 4 epochs to match the performance of the baseline trained with 8 epochs on WP. These results demonstrate that \aamtl~is both effective and efficient in terms of resource usage, time, and storage.

\vspace{1mm}\noindent\textbf{Ablation Study:} The balance between detection loss ($\mathcal{L}{D}$) and attribution loss($\mathcal{L}{A}$) is crucial for optimizing \aamtl. By default, we set their weighting factors ($\lambda_{D}$ and $\lambda_{A}$, as defined in Eq. 5) to 0.5 each. However, our framework allows users to adjust these weights based on their specific needs and task priorities. To analyze the impact of this weighting choice, we conduct an ablation study by varying $\lambda_{D}$ and $\lambda_{A}$. We evaluate RoBERTa on six datasets where \aamtl~achieves the most significant improvements: Essay, WP, and Reuters from MGTBench, \multitude, as well as arXiv and Reddit from M4.

The results are summarized in Table \ref{table:ablation}. Increasing the detection weight ($\lambda_{D} = 0.7, \lambda_{A} = 0.3, \lambda_{D}>\lambda_{A}$) slightly improves binary detection performance but significantly reduces attribution accuracy. This supports our observation that detection is generally easier than authorship attribution. Conversely, prioritizing attribution ($\lambda_{D} = 0.3, \lambda_{A} = 0.7, \lambda_{D}<\lambda_{A}$) enhances LLM differentiation but weakens detection performance.  Notably, on three of the six datasets (WP, Reuters, and Reddit), attribution performance with a higher $\lambda_{A}$ is even lower than our default setting. These findings suggest that attribution benefits from strong detection, and setting $\lambda_{D}=\lambda_{A}=0.5$ provides a balanced trade-off, ensuring high detection accuracy while maintaining reliable attribution performance.

\begin{table}[htbp]
\centering
\caption{Classification F1 scores (\%), number of parameters, and model sizes (GB) of \aamtl~and SOTA models. WP: WritingPrompt. RT: Reuters. MU: \multitude. PR: PeerRead. RD: Reddit. WH: Wikihow. Wiki: Wikipedia. Para.: Number of model parameters.} \vspace{-2mm}
\setlength{\tabcolsep}{1.6 pt}
\renewcommand{\arraystretch}{1.2}
\begin{tabular}{c|ccccccccc|cc}
\hline
\textbf{Model} & \textbf{Essay} & \textbf{WP} & \textbf{RT}& \textbf{MU} & \textbf{arXiv} &\textbf{PR} &\textbf{RD} &\textbf{WH}&\textbf{Wiki} &\textbf{Para.} & \textbf{Size} \\ \hline
\multicolumn{11}{c}{\textbf{ (a) Feature-based ML Classifiers}} \\ \hline
TF-IDF + MNB & 33.2 & 49.1 & 36.9 & 34.5 & 56.5 & 56.0 & 33.0 & 31.0 & 38.6 & - & 0.12 \\ \hline
TF-IDF + LR & 58.6 & 72.5 &  71.4 & 55.6  & 88.2 & 94.4 & 77.0 & 76.8 & 77.6 & - & 0.06 \\ \hline
BoW + SVM & 68.9 & 68.8 & 81.9 & 54.3  & 88.9 & 95.0 & 81.9 & 76.9 & 76.7 & - & 0.02 \\ \hline
BoW + RF & 66.3 & 62.8 & 76.7 & 42.6  & 82.1 & 94.3 & 76.0 & 75.9 & 78.6 & - & 0.05 \\ \hline\hline

\multicolumn{11}{c}{\textbf{(b) Fine-tuned Language Models}} \\ \hline
BERT & 88.6  & 86.3 & 90.2 & 69.1 & 95.7 & 98.8 & 87.7 & 94.0 & 90.0 & 110M &  0.42 \\ \hline
GPT-2 & 89.8 & 94.1 & 94.9& 74.9 & 97.5 & 99.3 & 93.4 & 96.3 & 92.5 & 124M &  0.55 \\ \hline
DistilBERT & 87.2 & 87.1 & 91.8 & 70.4  & 96.4 & 98.8 & 92.0 & 95.1 & 90.4 & 67M &  0.26 \\ \hline
mBERT &  90.4 & 89.9 & 92.8 & 70.2 & 94.8 & 99.0 & 91.4 & 94.8 & 91.7 & 178M &  0.68 \\ \hline
XLM-R & 87.3 & 87.2 & 92.1 & 70.6 & 95.2 & 98.8 & 91.8 & 94.7 & 90.0 & 278M &  1.08 \\\hline
RoBERTa & 95.7 & 94.0 & 95.2 & 73.3  & 94.4 & 99.0 & 93.2 & 96.0 & 92.7 & 125M &  0.48 \\ \hline
\hline

\multicolumn{11}{c}{\textbf{(c) \aamtl}} \\ \hline
DistilBERT & 91.6 & 89.5 & 93.6 & 72.8 & 98.2 & 99.4 & 92.9 & 95.9 & 91.9 & 67M &  0.26 \\ \hline
mBERT & 92.7 & 90.8 & 93.1 & 75.5 & 98.4 & 99.3 & 94.3 & 95.9 & 92.9 & 178M &  0.68\\ \hline
XLM-R & 90.3 & 90.7 & 93.2 & \textbf{77.1} & 98.1 & \textbf{99.4} & 92.6 & \textbf{96.9} & 93.3 & 278M &  1.08\\ \hline
RoBERTa & \textbf{96.6} & \textbf{95.6} & \textbf{96.6} & 75.9 & \textbf{98.4} & 99.3 & \textbf{95.4} & 96.5 & \textbf{93.3} & 125M &  0.48 \\ \hline
\end{tabular} 
\label{table:basic}
\end{table}

\begin{table}[H]
\centering
\caption{F1 scores (\%) of RoBERTa trained with \aamtl~and different $\lambda_D, \lambda_A$ weight values. WP: WritingPrompt. RT: Reuters. MU: \multitude. RD: Reddit. The highest F-1 score result for each dataset is marked in bold.}\label{table:ablation}
\setlength{\tabcolsep}{3 pt}
\renewcommand{\arraystretch}{1.2}
\begin{tabular}{c|cccccc|cccccc}
\hline
\multirow{2}{*}{$\lambda_D$ / $\lambda_A$} & \multicolumn{6}{c|}{$\mathcal{T}_{D}$} & \multicolumn{6}{c}{$\mathcal{T}_{A}$}\\  \cline{2-13}
& Essay & WP & RT & MU & arXiv & RD & Essay & WP & RT & MU & arXiv & RD \\
\hline
0.3 / 0.7 & 99.7 & 99.1 & 99.8 & 95.5 & 97.8 & 99.4 & \textbf{96.7} & 93.0 & 95.7 & \textbf{74.5} &  \textbf{99.0} & 92.7 \\
0.7 / 0.3 & 99.5 & 98.6 & 99.8 & \textbf{96.7} & 97.8 & \textbf{99.7} & 94.6 & 93.8 & 95.3 & 72.1 & 98.4 & 92.3\\
0.5 / 0.5 & \textbf{99.8} & \textbf{99.1} & \textbf{99.9} & 96.3 & \textbf{98.4} & 99.3 &  96.0 & \textbf{95.0} & \textbf{96.1} & 73.4 & 98.4 & \textbf{94.7} \\
\hline
\end{tabular}
\end{table}

\subsection{Cross-modal Proximity Analysis} \label{sec:crossmodel}
\begin{figure}[!htbp] 
    \centering
    \includegraphics[width=0.5\textwidth]{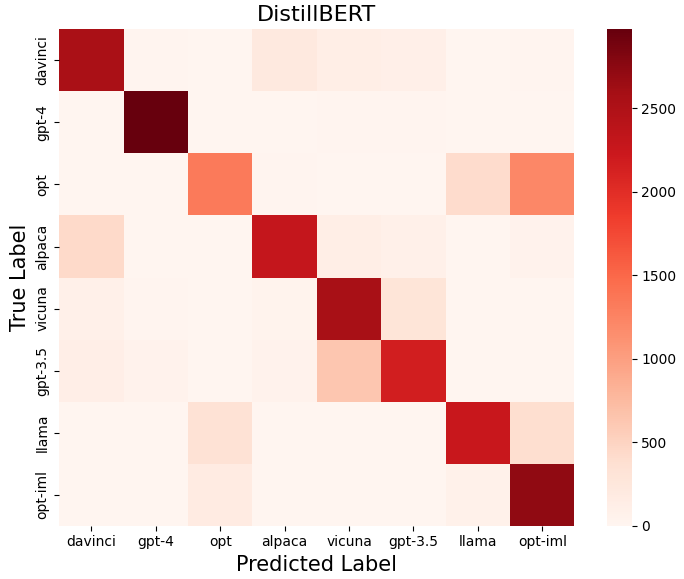}
    \includegraphics[width=0.48\textwidth]{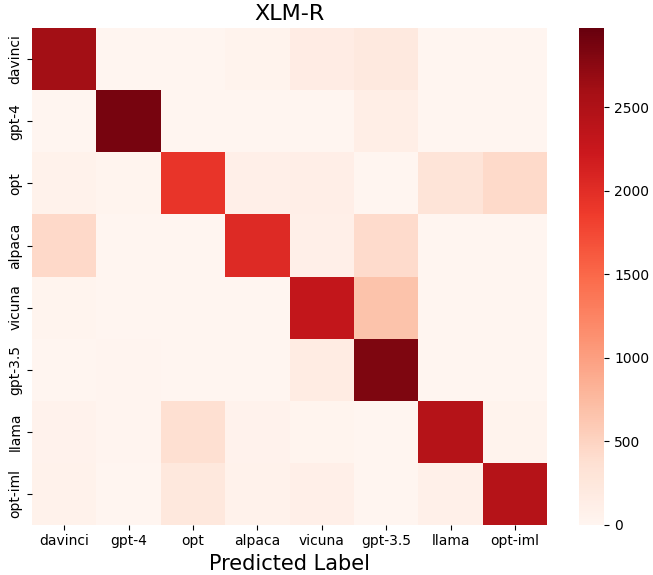}\vspace{-2mm}
    \caption{Confusion matrices of DistilBERT and XLM-R on \multitude.} 
    \label{fig:confusion1}
\end{figure}

To better understand LLM proximity and address \textbf{RQ2}, we examine the error patterns through the confusion matrices of \multitude, which is the most challenging dataset. We evaluate DistilBERT and XLM-R, which have the largest and smallest parameter sizes, respectively. As shown in Figure~\ref{fig:confusion1}, our analysis reveals consistent error patterns across classifiers: Alpaca-LoRA-30B is frequently misclassified as Text-DaVinci-003 and GPT-3.5, while Vicuna-13B and GPT-3.5 often confuse classifiers. Interestingly, relatively advanced models like OPT-66B are sometimes misclassified as smaller models, like OPT-IML-MAX-1.3B, with LLaMA-65B also being a common error. Conversely, smaller models, including OPT-IML-MAX-1.3B and those in the GPT family, are less likely to be misclassified as their more advanced counterparts.


The lineage of LLMs provides valuable insight into their similarities and differences: (1) \textbf{OPT and LLaMA ("OPT-like")}: Both families were developed by Meta (formerly Facebook AI Research). Their base models focus on pure next-token prediction without specific optimization for user interactions, making their responses less structured and user-friendly. Some variants of OPT models, such as OPT-IML and OPT-IML-Max, were fine-tuned by instruction following, but they still retain much of OPT-66B's behavior patterns. Both OPT and LLaMA adopt a decoder-only transformer-based architecture, while LLaMA achieves higher training efficiency and lower perplexity through better stability and data curation. Considering data, both of them leverage large, curated datasets from web crawls, academic corpora, and structured sources like Wikipedia, Reddit, and GitHub \cite{touvron2023llama1,zhang2022opt}. (2) \textbf{Alpaca-LoRA, GPT-3.5s, and Vicuna ("Davinci-like"):} Both Text-Davinci-003 and GPT-3.5 belong to OpenAI's InstructGPT series. Although their exact architectures and hyperparameters remain proprietary, they were known to be trained with large-scale Reinforcement Learning from Human Feedback (RLHF) \cite{rlhf}, which improves their ability to follow instructions, generate clear responses, and avoid harmful content. Community-driven models like Alpaca and Vicuna are built on them: Alpaca (e.g., Alpaca-LoRA-30B) was fine-tuned using Text-Davinci-003 outputs, and Vicuna (e.g., Vicuna-13B) was trained on ShareGPT dialogues, which often contain ChatGPT responses. As a result, these models inherit the stylistic, formatting, and instructional traits of OpenAI's models. (3) \textbf{GPT-4}: GPT-4 is a more advanced LLM with significant breakthroughs and distinctness. It excels at handling complex tasks with stronger reasoning, coherence, and problem-solving abilities than its predecessors.

\begin{tcolorbox}[title=Takeaway II]
\aamtl's misclassification patterns reveal relationships between LLMs, which often align with their lineage. 
\end{tcolorbox}

\begin{figure}[!htbp]
    \centering
    \includegraphics[width=1.0\textwidth, trim=0 0 290 0, clip]{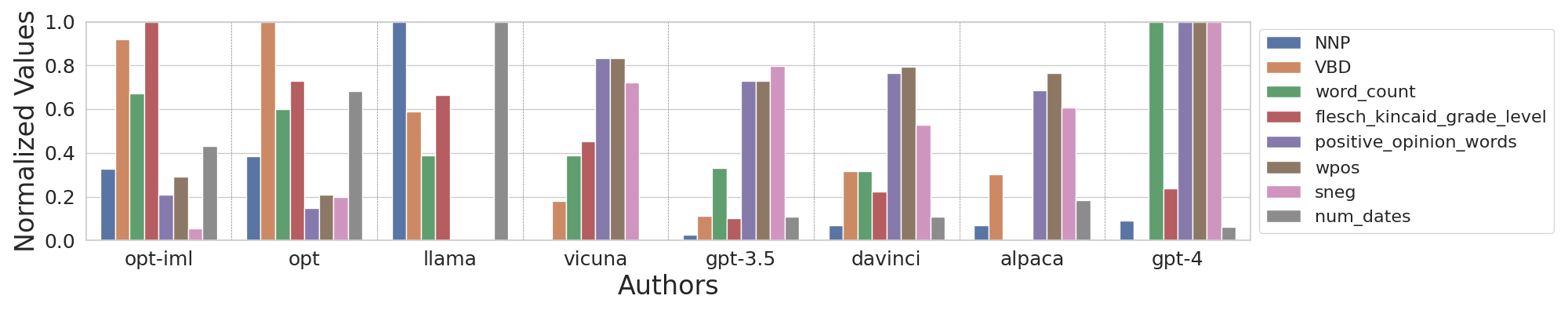}\vspace{-2mm}
    \caption{The average normalized values of the eight most important NELA features are shown. The features are listed in order of importance, from left (blue) to right (gray): NNP, VBD, total word count, Flesch-Kincaid readability score, positive opinion word count, wpos, sneg, and explicit date count.}
    \label{fig:nela_features}
\end{figure}

\vspace{0mm}\noindent\textbf{Stylometric Analysis.}
To better understand the proximities of the LLMs, we analyze their writing styles using NELA \cite{horne2018assessing}, a framework that evaluates text based on 87 features across six categories: style, complexity, bias, affect, morality, and events. From each LLM-generated sample, we extracted a NELA feature vector. We identified the eight most important features using the Chi-squared test: singular proper nouns (NNP) count, past tense verbs (VBD) count, total word count, Flesch-Kincaid readability score, positive opinion word count, Word-level Positive Sentiment Score (wpos),  Sentence-level Negative Sentiment Score (sneg), and explicit date mentions. Figure~\ref{fig:nela_features} illustrates the average feature values of LLM-generated text in \multitude~dataset.

Our analysis reveals that OPT-66B and OPT-IML-MAX-1.3B are the most similar, while Vicuna-13B, GPT-3.5, and Text-Davinci-003 form another closely related group. However, notable intra-group differences exist among the "OPT-like" models, "Davinci-like" models, and GPT-4. These align with our observations derived from our \aamtl~model's confusion matrices. Specifically, we have the following findings:

\begin{itemize}
\item \textbf{OPT and LLaMA:}
OPT models frequently use singular proper nouns, past tense verbs, and explicit dates, while showing minimal sentiment word usage. Their responses are lengthy, though shorter than the ones generated by GPT-4, and are complex and harder to read. LLaMA-65B shares these patterns but with an even stronger preference for proper nouns and explicit dates while avoiding sentiment expressions. LLaMA's responses are slightly easier to read than OPT's but still more complex than those of the other LLMs.

\item \textbf{Vicuna, GPT-3.5, Text-Davinci-003, and Alpaca:}
These models exhibit highly similar patterns: they use far fewer singular proper nouns, past tense verbs, and explicit dates, which is nearly half as frequently as OPT models. Instead, they incorporate significantly more sentiment and opinion words, which is almost 50\% more than OPT and LLaMA. They generate shorter and more readable responses. Alpaca follows this trend but stands out with the shortest responses (under 150 words on average) and the highest readability.

\item \textbf{GPT-4:}
GPT-4 shares some similarities with "Davinci-like" models but stands out with significantly longer responses, averaging nearly 300 words. Despite this, its readability remains low, comparable to Text-Davinci-003. Additionally, GPT-4 uses more sentiment and opinion words than Vicuna, GPT-3.5, Text-Davinci-003, and Alpaca.
\end{itemize}

However, while NELA feature analysis provides valuable insights into LLM behaviors, it does not fully explain their differences. To validate this, we trained two ML models, SVM and LR, using NELA features as text embeddings. Both models performed poorly, with F1 scores of only 41.9\% (SVM) and 31.6\% (LR). This suggests that LLM stylistic differences involve more subtle factors beyond what NELA features can represent.


\subsection{Cross-Lingual Analysis} \label{sec:crosslingual}
\noindent\textbf{Multi-Lingual Text Detection and Attribution.}
We use mBERT and XLM-R for multi-lingual tasks. To better analyze the impact of different languages, we exclude highly correlated and potentially confusing models (e.g., OPT-66B versus OPT-IML-MAX-1.3B) and focus on five representative models: OPT-IML-MAX-1.3B, LLaMA-65B, GPT-4, Text-DaVinci-003, and Vicuna-13B. As shown in Table~\ref{tab:model_comparison_by_language}, the classifier's F1 scores improve for both Task 1 and 2 after applying \aamtl. Across nine languages, mBERT gains an average increase of 2.96\% for Task 1 and 2.67\% for Task 2, while XLM-R shows improvements of 1.78\% and 2.6\%, respectively. These results confirm that the two tasks complement each other. Among the languages, Dutch (NL) and Portuguese (PT) proved to be the most challenging for both tasks. However, incorporating \aamtl~significantly improved detection F1 scores for these two languages, reaching F-1 scores of 99.83\% and 98.17\%, respectively. Moreover, no signs of overfitting are shown for either task when applying \aamtl. This again aligns with our hypothesis that \aamtl~serves as a form of regularization to prevent overfitting to single tasks.

\begin{tcolorbox}[title=Takeaway III]
\aamtl~significantly improves multilingual detection and attribution, especially in challenging languages.
\end{tcolorbox}

\begin{table}[!htbp]
\centering
\caption{F1 scores (\%) of mBERT and XLM-R w/o and with \aamtl~by language. \blue{Blue} and \red{red} numbers show the differences. Gains larger than $3\%$ are in bold.}\vspace{-2mm}
\setlength{\tabcolsep}{2.5 pt}
\renewcommand{\arraystretch}{1.2}
\begin{tabular}{c|c|ccccccccc}
\hline
\textbf{Tasks} & \textbf{Models} & \textbf{EN} & \textbf{ES} & \textbf{RU} & \textbf{UK} & \textbf{CS} & \textbf{DE} & \textbf{NL} & \textbf{CA} & \textbf{PT} \\ \hline

\multirow{4}{*}{$\mathcal{T}_D$} & mBERT & 98.02 & 97.93 & 96.33 & 96.49 & 98.45 & 96.24 & 96.13 & 97.93 & 95.17 \\
& \myplus MTL & \blue{(\myplus 2.5)} & \blue{(\myplus 1.5)} & \blue{(\myplus 0.5)} & \blue{(\myplus 1.6)} & \blue{(\myplus \textbf{3.1})} & \blue{(\myplus \textbf{3.9})} & \blue{(\myplus \textbf{3.7})} & \blue{(\myplus 2.9)} & \blue{(\myplus \textbf{3.0})} \\\cline{2-11}

& XLM-R & 97.25 &  97.87 &  98.80 &  96.34 &  96.19 &  96.87 &  95.53 &  97.76 &  95.36\\ 
& \myplus MTL & \blue{(\myplus 1.5)} & \blue{(\myplus 1.5)} & \blue{(\myplus 2.3)} & \blue{(\myplus 1.6)} & \blue{(\myplus 0.6)} & \blue{(\myplus 2.1)} & \blue{(\myplus 2.0)} & \blue{(\myplus 2.0)} & \blue{(\myplus 2.4)} \\\hline

\multirow{4}{*}{$\mathcal{T}_A$} & mBERT & 91.45 & 94.60 & 95.73 & 91.20 & 91.43 & 89.39 & 87.60 & 90.03 & 87.69 \\
& \myplus MTL & \blue{(\myplus 2.9)} & \blue{(\myplus \textbf{4.6})} & \blue{(\myplus \textbf{3.5})} & \blue{(\myplus \textbf{6.7})} & \blue{(\myplus 2.5)} & \blue{(\myplus 1.8)} & \red{(\myminus 0.1)} & \red{(\myminus 0.3)} & \blue{(\myplus 2.4)} \\\cline{2-11}

& XLM-R &  91.36 &  93.84 &  96.58 &  90.92 &  91.21 &  87.80 &  88.76 &  90.89 &  82.54\\
& \myplus MTL & \blue{(\myplus 2.1)} & \blue{(\myplus \textbf{3.4})} & \blue{(\myplus \textbf{3.5})} & \blue{(\myplus \textbf{3.7})} & \blue{(\myplus \textbf{3.9})} & \blue{(\myplus 1.0)} & \blue{(\myplus \textbf{4.9})} & \red{(\myminus 0.4)} & \blue{(\myplus 1.3)} \\\hline

\end{tabular}
\label{tab:model_comparison_by_language}
\end{table}

\begin{figure}[!htbp] 
    \centering
    \includegraphics[width=0.495\textwidth]{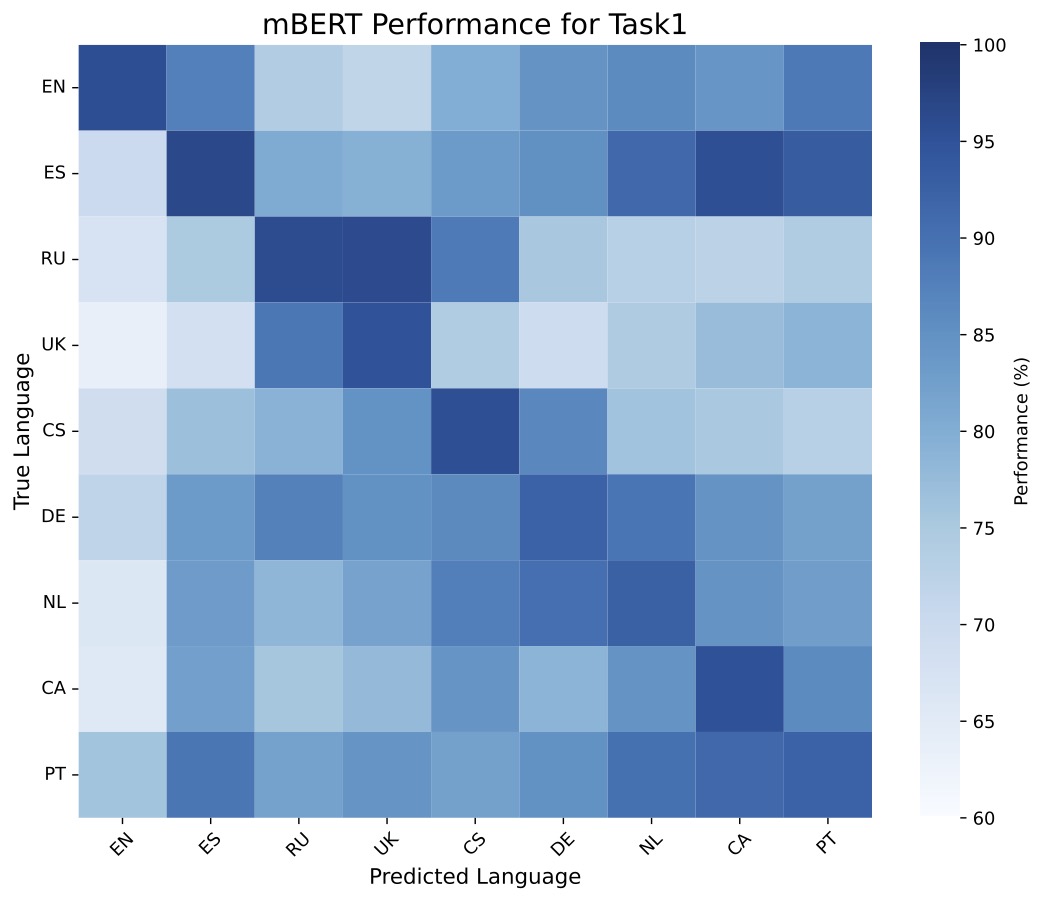}
    \includegraphics[width=0.495\textwidth]{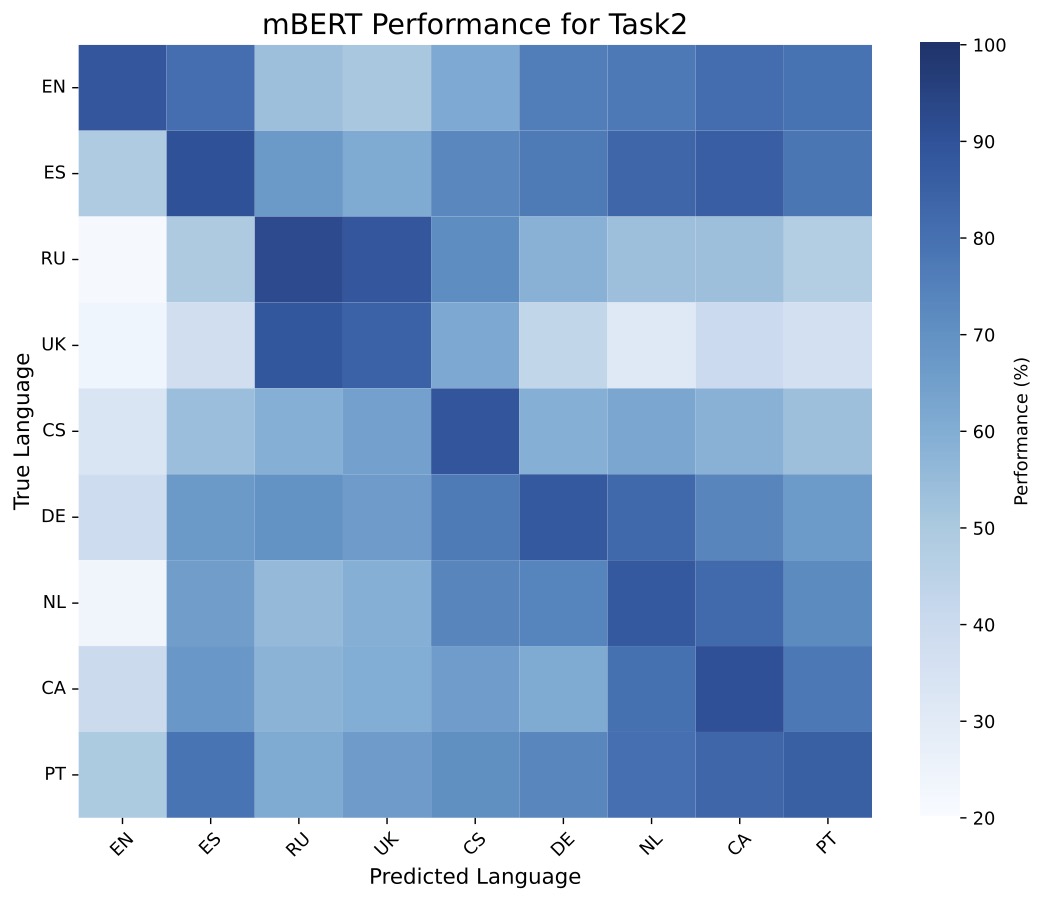}
    \includegraphics[width=0.495\textwidth]{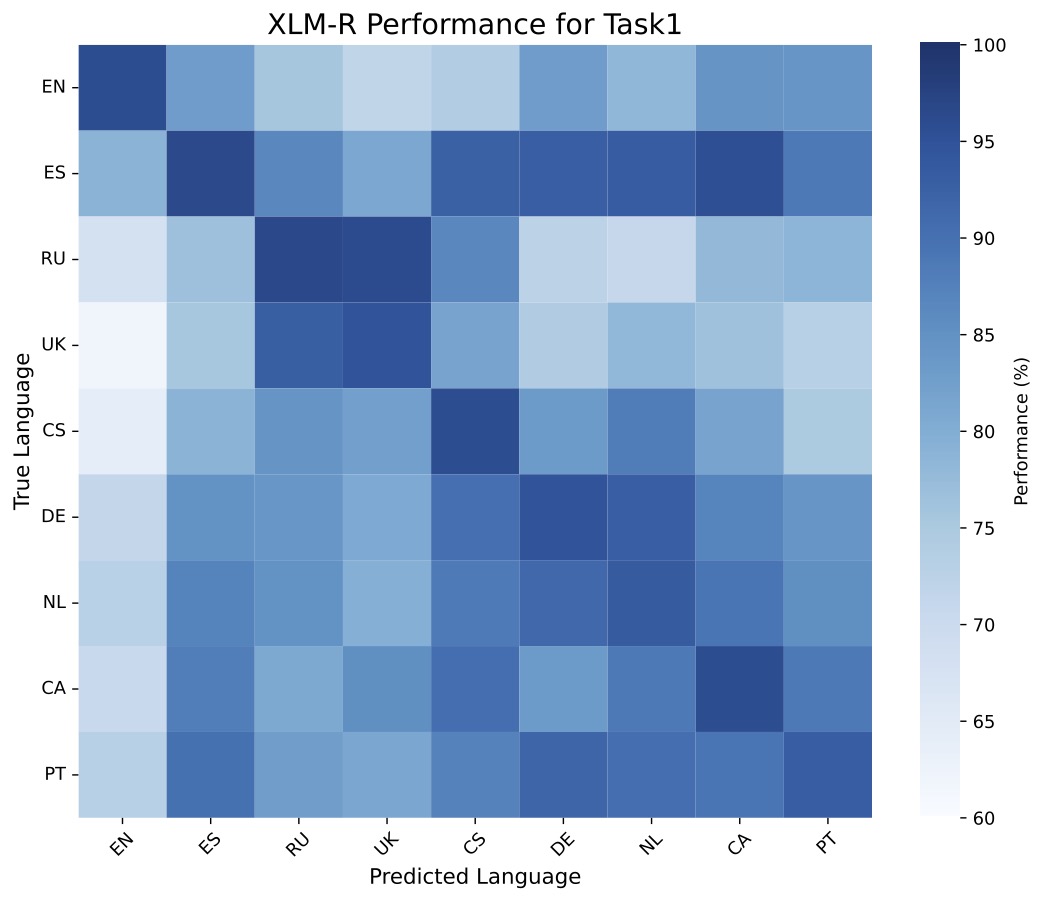}
    \includegraphics[width=0.495\textwidth]{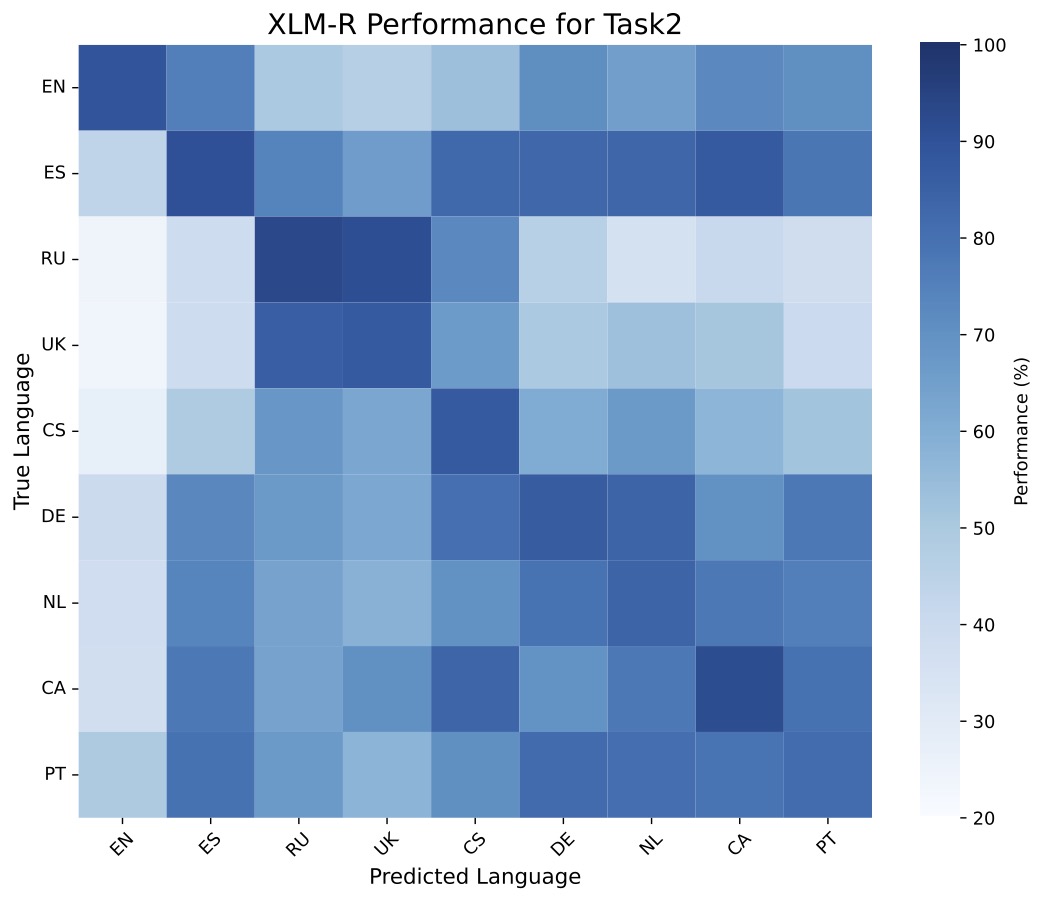}
    \vspace{-2mm}
    \caption{Cross-Lingual validations of models fine-tuned on \multitude. From left to right: mBERT for Task1 \& Task 2 and XLM-R for Task 1 \& Task 2.} 
    \label{fig:confusion2}
\end{figure}

\noindent\textbf{Cross-Lingual Generalization.}
We follow \cite{macko2023multitude} to evaluate cross-lingual generalization, addressing \textbf{RQ3}. Figure~\ref{fig:confusion2} presents the cross-validation performance of mBERT and XLM-R on Tasks 1 and 2.

   Each model was fine-tuned on individual source languages, and results show that they can generalize to other languages. However, their performance is generally lower than when fine-tuned directly on the target languages. Notably, XLM-R fine-tuned on Ukrainian transfers well to Russian and vice versa, demonstrating strong cross-lingual performance among closely related Slavic languages that share the Cyrillic script. In Task 2, performance varies widely, with mBERT achieving accuracies between 21.31\% and 88.7\%, while XLM-R ranges from 23.58\% to 91.07\%.

Further analysis reveals that language families and writing systems significantly impact cross-lingual transfer. For instance, models generalize well among Romance languages like Spanish, Portuguese, and Catalan. Most source languages transfer effectively to Czech, which uses a Latin script. Fine-tuning on English or Spanish typically results in strong generalization across multiple languages, whereas models fine-tuned on Ukrainian or Russian exhibit poor transferability except to Czech. Interestingly, models trained on English show moderate performance on other Germanic languages, particularly Dutch and German, likely due to English's historical divergence and influence from Latin-based languages.

\subsection{Robustness} \label{sec:robust}
We assess the robustness of \aamtl~against two widely used authorship obfuscation techniques:

\begin{itemize}
\item \textbf{Mutant-X (MX):} A stylometric obfuscation method that uses genetic algorithms \cite{mahmood2019mutantx}. It alters text by replacing words with synonyms and rearranging phrases while maintaining semantic similarity. These modifications are guided by a fitness function that balances authorship attribution probability with meaning preservation. We use the publicly available implementation with its default settings \footnote{https://github.com/asad1996172/Mutant-X}.

\item \textbf{Back-Translation (BT):} A technique that translates text into another language and then back to the original \cite{keswani2016author,almishari2014fighting}. This process changes word choice and sentence structure while preserving meaning. For our experiments, we use French as the intermediate language, utilizing Opus MT models for English-to-French ("en-fr") and French-to-English ("fr-en") translation \cite{tiedemann2023democratizing,TiedemannThottingal}.
\end{itemize}

Table~\ref{tab:model_robust} shows how applying these obfuscation techniques affects the accuracy of our XLM-R classifier trained with \aamtl. We conduct this evaluation using 500 randomly selected English text samples from \multitude. Our results indicate that while the model remains generally robust, the extent of accuracy degradation varies across LLMs. For Mutant-X, the classifier is most resistant when analyzing text from OPT-66B, where accuracy decreases by only 0.61\%, suggesting that text generated by OPT-66B has fewer distinct stylistic features, making it harder to manipulate. In contrast, text from GPT-4 is the most affected, with accuracy dropping by 9.03\%, indicating that its unique vocabulary and structured writing style make it more vulnerable to synonym substitutions and word-order changes. For Back-Translation, the classifier performs best on text from Vicuna-13B, with only a 4.38\% accuracy drop. However, it struggles significantly with text generated by GPT-3.5, Text-DaVinci-003, and LLaMA-65B, where accuracy declines by 14\% to 17\%. Interestingly, we find a negative correlation between the model's initial classification accuracy and its resilience to Back-Translation. Texts that were harder to classify originally suffer the most after translation, likely because the process disrupts subtle linguistic patterns.

\begin{table}[!htbp]
\centering
\caption{Accuracy differences (\%) against authorship obfuscation. N/A: Original accuracy on original text. MX: Mutant-X. BT: Back-Translation.}\vspace{-2mm}
\setlength{\tabcolsep}{2.5 pt}
\renewcommand{\arraystretch}{1.2}
\begin{tabular}{c|cccccccc}
\hline
{\textbf{Atk.}} & \textbf{GPT-3.5} & \textbf{GPT-4} & \textbf{DaVinci} & \textbf{Vicuna} & \textbf{OPT} & \textbf{OPT-IML} & \textbf{Alpaca} & \textbf{LLaMA} \\ \hline
{N/A} &86.43  &91.70  & 59.57 & 91.75 & 52.18 & 91.31 & 67.29 & 57.40 \\ \hline\hline
{MX} & \myminus 2.68  & \myminus 9.03  &\myminus 2.53  & \myminus 4.75 &\myminus 0.61  & \myminus 2.86 &\myminus 2.67  & \myminus 5.22 \\ \hline
{BT} & \myminus 14.2 & \myminus 8.31  &\myminus 14.5 &\myminus 4.38  & \myminus 10.4 &\myminus 8.64  &\myminus 14.9  &\myminus 17.0 \\
\hline
\end{tabular}
\label{tab:model_robust}
\end{table}

These findings demonstrate that the robustness of \aamtl~varies depending on the LLM source of the analyzed text. While some LLM-generated texts are naturally more resistant to stylometric obfuscation, others are more susceptible, making it easier for adversaries to evade attribution. Although \aamtl~exhibits strong real-world applicability against mild obfuscation techniques, it is not explicitly designed to maximize robustness. We do not claim that \aamtl~offers superior resistance to adversarial attacks. In future work, we aim to explore adaptive defenses that dynamically adjust to different obfuscation strategies and develop more resilient authorship attribution methods that leverage deeper linguistic and structural features.

\section{Conclusion}  \label{sec:conclusion}
In this paper, we introduce \aamtl, a multi-task learning framework designed to detect and attribute LLM-generated text jointly. We highlight the benefits of multi-task learning by evaluating on nine datasets and with four different backbone models. The results show that multi-task learning improves performance for both tasks and in both English and multi-lingual contexts. Using our accurate models, we further analyze cross-modal and cross-lingual similarities and evaluate \aamtl~against various authorship obfuscation techniques. Our work offers an effective tool as well as valuable insight into LLM behaviors and proximity, contributing to a deeper understanding of AI-generated content and its associated security and safety risks.

\bibliographystyle{splncs04}
\bibliography{ref}

\end{document}